\documentclass[11pt]{iopart} % IOP journal style

\usepackage{graphicx}    % Include figure files
\usepackage{latexsym}    % 
\usepackage{color}       % 
\usepackage{dcolumn}     % Align table columns on decimal point
\usepackage{bm}          % bold math
\usepackage{amssymb}     % for math
\usepackage{hyperref}    % add hypertext capabilities

\usepackage{graphics}
\usepackage{epsfig}
\usepackage{caption}
\begin{document}

\title{Structures and Lagrangian statistics of the Taylor-Green Dynamo}

\author{H Homann$^1$, Y Ponty$^1$, G Krstulovic$^1$, R Grauer$^2$}
\address{$^1$ Universit\'e de Nice-Sophia, CNRS, Observatoire de la C\^ote d'Azur, 
CS 34229, 06304 Nice Cedex 4, France}
\address{$^2$ Institut f\"ur Theoretische Physik I, Ruhr-Universit\"at Bochum, 44780 Bochum, Germany}

\begin{abstract}
  The evolution of a Taylor-Green forced magnetohydrodynamic (MHD)
  system showing dynamo activity is analyzed via direct numerical
  simulations. The statistical properties of the velocity and magnetic
  field in Eulerian coordinates and along trajectories of fluid
  elements (Lagrangian coordinates) are found to change between the
  kinematic, non-linear and saturated regime.  Fluid element (tracer)
  trajectories change from chaotic quasi-isotropic (kinematic phase)
  to mean magnetic field aligned (saturated phase). The probability
  density functions (PDFs) of the magnetic field change from strongly
  non-Gaussian in the kinematic to quasi-Gaussian PDFs in the
  saturated regime so that their flatness give a precise handle on the
  definition of the limiting points of the three regimes. Also the
  statistics of the fluctuations of the kinetic and magnetic energy
  along fluid trajectories change. All this goes along with a
  dramatic increase of the correlation time of velocity and magnetic
  field fluctuations experienced by tracers significantly exceeding one
  turbulent large-eddy turn-over time. A remarkable consequence is an
  intermittent scaling regime of the Lagrangian magnetic field
  structure functions at unusually long time scales.
\end{abstract}

%\pacs{52.30.-q, 52.65.-y, 52.30.Cv}
% 52.30.-q Plasma dynamics and flow
% 52.65.-y Plasma simulation
% 52.30.Ex Two-fluid and multi-fluid plasmas
% 52.30.Cv Magnetohydrodynamics (including electron magnetohydrodynamics)
% 52.35.Mw Nonlinear phenomena: waves, wave propagation, and other interactions
% (including parametric effects, mode coupling, ponderomotive effects, etc.) 

\maketitle
%\let\clearpage\relax
%\nopagebreak
\section{introduction}
The magnetic field of stars and telluric planets is explained by the
dynamo instability, produced by a turbulent conducting fluid where the
induction due to the motion takes over the magnetic diffusion.  A
system with dynamo action passes through different stages from the
linear (kinematic) to the non-linear and finally to the saturated
phase. During the kinematic phase the magnetic energy grows
exponentially but has no influence on the flow. During the non-linear
phase the Lorentz force changes the flow and at the saturation state
the diffusion and the electromotive force reach an equilibrium. The
system reaches the fully non-linear magnetohydrodynamic (MHD) regime,
driven by an energy exchange between the velocity and magnetic field.

In the last decade, several experimental groups have investigated
dynamo action in laboratory experiments using liquid
sodium~\cite{gailitis2000, gailitis2001, muller2000,
  stieglitz2001,monchaux2007,berhanu2007}.  The instability threshold
and a rich non-linear behavior along the saturation regime have been
largely observed
\cite{monchaux_von_2009,berhanu_bistability_2009,berhanu_dynamo_2010,gallet_experimental_2012}.
Investigations of the fast kinematic dynamo \cite{STFbook} in the
90's, showed the important role of the chaotic properties of the fluid
trajectories for the dynamo threshold. The effect of turbulent
fluctuations on the dynamo onset has also been studied in
Navier-Stokes flows \cite{ponty2007, ponty2011} and some noisy models
\cite{dubrulle_bifurcations_2007}, distinguishing the important role
between the mean flow dynamo and fluctuation dynamo modes
\cite{Schekochihin2007}.

The transition between the linear and the saturation regime has been
studied measuring the finite-time Lyapunov exponent of the flow
trajectories \cite{cattaneo1996, zienicke1998}. These results showed a
strong reduction of the chaotic properties in the saturation phase,
due to the action of the Lorentz force. Fluid trajectories thus change
along the different dynamo phases. This naturally motivates the use of
a Lagrangian description where the properties of the flow are
investigated by using tracers \cite{JFPcatania}.

Lagrangian statistics describe the dynamical evolution of physical
quantities along tracer trajectories in contrast to the Eulerian
perspective in which such quantities are analyzed on fixed spatial
points. During the last decades Lagrangian studies revealed new
aspects of homogeneous and isotropic
turbulence~\cite{toschi-bodenschatz-2009}. The trajectory point of
view has shown to be especially useful for the study of coherent
structures and intermittency \cite{biferale:2004b,
  yeung-pope-etal:2006, bec-etal:2006} as well as for diffusion and
dispersion properties of hydrodynamic and homogeneous and isotropic
MHD turbulence
\cite{yeung-borgas:2004,bitane-homann-etal-2013,eyink2013}. This
approach has produced many experimental results by tracking solid
inertial particles or bubbles
\cite{mordant_measurement_2001,mordant_long_2002,la_porta_using_2000,la_porta_fluid_2001},
which can have a different dynamics compared to simple
tracers. Lagrangian statistics has also been used in MHD simulations
\cite{homann_lagrangian_2007, homann_bridging_2009} to compare the
anomalous exponents of the structure function to their hydrodynamic
counterparts and to understand the relation between Eulerian and
Lagrangian statistics.

In this study we use a Lagrangian approach in the context of turbulent
dynamo action. In the Kraichnan-Kazantsev framework it is used to
determine the dynamo onset as a function of the roughness of the
flow~\cite{Celani2006}. In this work the Lagrangian perspective allows
us to precisely define the limiting points of each regime, to
highlight the correlation of the magnetic field and the fluid
trajectories and to discover a to our knowledge unknown scaling regime
at time scales well beyond the large-eddy turn-over time of the
turbulent flow.

The dynamo systems under consideration in this work is induced mainly
by the so called Taylor-Green forcing~\cite{brachet1983}. For
comparison, a second forcing where the large scales of the velocity
field are frozen in time is also considered. The Taylor-Green flow,
which can produce dynamo action, is a very well documented study case
involving rich dynamo behaviors in both the linear and the non-linear
regime \cite{nore,nore2,ponty2005,
  mininni2005_nl,ponty2007,dubrulle_bifurcations_2007,ponty2008,giorgio2011}.
The Taylor-Green flow contained in a periodic box has several
properties that mimic the von K\`arm\`an flow, driven by two
counter-rotating impellers. The von K\`arm\`an flow has a strong
experimental history inside the turbulent and the dynamo scientific
communities.  Indeed, several teams set up such experiments with
different designs in incompressible
\cite{dijkstra_flow_1983,douady_direct_1991} and compressible flows
\cite{fauve_pressure_1993,pinton_1994, abry_analysis_1994,
  labbe_study_1996, cadot_energy_1997, de_la_torre_slow_2007}. This
type of experiment was also one of the first setup used to study
Lagrangian statistics of turbulent flows
\cite{mordant_measurement_2001,mordant_long_2002,la_porta_using_2000,la_porta_fluid_2001}. After
the intensive water campaign experiments of the Saclay group
\cite{marie_these_2003, marie2004, ravelet_multistability_2004,
  ravelet_these_2005,ravelet_supercritical_2008}, this setup has lead
to the von K\`arm\`an Sodium (VKS) dynamo campaigns also cited above.
 
This paper is organized as follows: In the next section the dynamo
system under consideration and the numerical method are explained. In
section~\ref{sec:dynAc} the evolution of the flow structure and
trajectories within the three different regimes are discussed. Section
\ref{sec:magFluc} investigates the changes of the velocity and
magnetic field fluctuations from one regime to another. A magnetic
field scaling regime at very large temporal scales is reported in
section ~\ref{sec:longTime}. Conclusions are drawn in section
~\ref{sec:conclusion}.

\section{Model and methods}

We perform direct numerical simulations of turbulent MHD flows with
large-scale forcing in a periodic box. We integrate the
three-dimensional incompressible MHD equations that, expressed in
Alfv\'enic units, read
\begin{eqnarray}
  \label{momentum}
  && \partial_t {\bm u} + \bm u\cdot\nabla\bm u = (\nabla\times{\bm
    B}) \times {\bm B} - \nabla p +\nu \nabla^2 {\bm u} + \bm F
  ,\ \ \ \strut \label{eq:mhd1} \\ && \partial_t {\bm B} =
  \nabla\times({\bm u}\times{\bm B})+\eta \nabla^2 {\bm
    B},\label{eq:mhd2}\\ && \nabla\cdot {\bm u} = 0, \quad \nabla
  \cdot {\bm B} = 0,\label{eq:mhd3}
\end{eqnarray}
where $\nu$ and $\eta$ are the kinematic viscosity and the magnetic
diffusivity respectively. The density of the fluid is set to unity and
$\bm F$ is the constant volume force which generates and maintains the
turbulent flow.  For most simulations (see table \ref{table1}) we
consider the so called Taylor-Green (TG) flow that is generated by
forcing with the TG vortex
\begin{equation}
\label{eq:tg}
\bm{F}_{\rm TG}=F_0\left(\, \sin(k_0x)\cos(k_0y)\cos(k_0z) \,, \,\cos(k_0x)\sin(k_0y)\cos(k_0z)\,, \,0 \right)
\end{equation}
with $F_0=3$ and $k_0=1$ or $k_0=2$. When $k_0=1$, this forcing leads
to a subdivision of the total domain in eight fundamental boxes that,
when symmetries are preserved, can be related to each other by
mirror-symmetric transformations. Each fundamental box contains a
swirling flow composed of a shear layer between two counter-rotating
eddies. The TG flow mimics some aspects of the von K\`arm\`an flow,
largely used in hydrodynamics turbulence and dynamo action
experiments. As we will see in the following sections, the TG dynamo
action presents some particular properties due to its anisotropy. In
order to distinguish universal and non-universal properties of the TG
dynamo action we also consider a mechanical force ${\bf F}$ (see run
ffH and ff in table \ref{table1}) obtained by keeping constant
all Fourier modes of the velocity field in the two lowest shells
(henceforth called frozen force simulation (FF)). This force is much
less structured than the TG forcing and the corresponding flow is
found to be nearly isotropic.

\begin{table}
\footnotesize
\centering
\begin{tabular}{ccccccc}
    run & regime     & $Pm$ & $Re$ & force type      & $N$  & $N_p$\\
    \hline 
    tgH   & hydro       &   -  & 350  & TG ($k_0=1$)  & $256^3$ & $2\cdot 10^5$ \\
    tg1 & saturated   & 1/4  & 297  & TG ($k_0=1$)    & $256^3$ & $2\cdot 10^5$ \\
    tg2 & saturated   & 1/2  & 670  & TG ($k_0=1$)    & $256^3$ & $2\cdot 10^5$ \\
    tg3 & saturated   & 1    & 676  & TG ($k_0=1$)    & $256^3$ & $2\cdot 10^5$ \\
    tg4 & saturated   & 1    & 128  & TG ($k_0=2$)    & $256^3$ & $2\cdot 10^5$ \\
    tg5 & saturated   & 1    & 545  & TG ($k_0=1$)    & $128^3$ & $2\cdot 10^4$ \\
    ffH & hydro       & -    & 476  & FF (frozen)     & $256^3$ & $2\cdot 10^5$\\
    ff  & saturated   & 1    & 625  & FF (frozen)     & $256^3$ & $2\cdot 10^5$
\end{tabular}
\caption{\label{table1}List of the numerical
  simulations. $Pm=\nu/\eta$: magnetic Prandtl number, $Re =
  u^\mathrm{rms}\,L/ \nu$: Reynolds number, $u^\mathrm{rms}$ root mean
  square velocity (defined in table~\ref{table2}), $N$ number of
  collocation points, $N_p$ number of tracer particles.}
\end{table}

We use classical global quantities to characterize the different
dynamo phases. The kinetic energy $E_{\rm kin}$, the magnetic energy
$E_{\rm mag}$, the enstrophy $\Omega$, the cross helicity $H_C$, and
the magnetic helicity $H_M$ are defined as
\begin{eqnarray*}
  &E_{\rm kin}=\frac{1}{2}\left<{\bf u}^2\right>\,,\,E_{\rm mag}=\frac{1}{2}\left<{\bf B}^2\right>\,,\,\Omega=\frac{1}{2}\langle(\nabla \times \bm{u})^2\rangle\,,\\
  &H_C=\left<{\bf u} \cdot {\bf B}\right>\,,\,H_M=\left<{\bf A} \cdot {\bf B}\right>,
\end{eqnarray*}
where $\langle$ $\rangle$ stands for spatial average and ${\bf B} =
\nabla \times {\bf A}$ with ${\bf A}$ the magnetic potential. In the
ideal case ($\nu=\eta=0$) and without forcing (${\bf F}=0$), $E_{\rm
  tot}=E_{\rm kin}+E_{\rm mag}$, $H_C$ and $H_M$ are conserved by
the MHD equations. 

The TG forcing (\ref{eq:tg}) is purely horizontal, leading to
anisotropy in both the hydrodynamic and the MHD regime. In order to
quantify this anisotropy we compute the root mean square (rms) values
of the perpendicular ($xy$-plane) and parallel ($z$) components of the
velocity fields and define the global isotropy coefficient
$\rho_u^{iso}$ as
\begin{eqnarray*}
&u_{\perp}^{\rm rms}=[\langle u_x^2+u_y^2\rangle/ 2\,]^{1/2}\,,\,u_{\parallel}^{\rm rms}=\langle u_z^2\rangle^{1/2}\,,\\
&\rho^{iso}_{u}=u_{\parallel}^{\rm rms}/u_{\perp}^{\rm rms},\,\rho^{iso}_{B}=B_{\parallel}^{\rm rms}/B_{\perp}^{\rm rms}.
\end{eqnarray*}
With these definitions the average rms velocity is
$u^\mathrm{rms}=[(2\,(u_\perp^{\mathrm{rms}})^2+(u_\parallel^\mathrm{rms})^2)/3]^{1/2}$. We
use analog definitions for the magnetic field.

There are also three important dimensionless numbers, namely the
kinematic Reynolds number $Re$, the magnetic Reynolds number $Rm$ and
the magnetic Prandtl number $Pm$ defined as
\begin{equation}
Re=\frac{L\,u^{\rm rms}}{\nu},\hspace{.5cm} Rm=\frac{L\, u^{\rm rms}}{\eta},\hspace{.5cm} Pm=\frac{Rm}{Re}=\frac{\nu}{\eta}\label{Eq:defReynolds}.
\end{equation}
A useful quantity that measures the alignment of velocity and magnetic
fields is the normalized cross helicity defined as $h_C= \cos{[\bf u},
  {\bf B}]=H_C/(u^\mathrm{rms}\,B^\mathrm{rms})$.

Numerical integration of the MHD equations
(\ref{momentum}-\ref{eq:mhd3}) is carried out by using a fully
MPI-parallel pseudo-spectral code (LaTu
\cite{homann-schulz-etal:2011}) with a strongly stable third order
Runge-Kutta temporal scheme. De-aliasing is performed using the
standard $2/3$ rule.

In this work we are concerned with the Lagrangian aspects of dynamo
action.  Lagrangian statistics are obtained by tracers obeying the
equations
\begin{equation}
  \label{Eq:tracers}
  \dot{\bm X}({\bm x}, t) = {\bm v}(t), \quad {\bm v}(t) = {\bm u}({\bm X}({\bm x},t)),
\end{equation}
where ${\bm X}({\bm x}, t)$ and ${\bm v}(t)$ are the position and
velocity of a tracer which started ${\bm x}$ for $t=0$. The magnetic
field along a tracer path is denoted by ${\bm b}(t) = {\bm B}({\bm
  X}({\bm x},t))$. The values of the velocity fields and other
physical quantities at the particle positions are evaluated and stored
using a tricubic interpolation which is numerically efficient and
accurate~\cite{homann-dreher-etal:2007}. The equation
(\ref{Eq:tracers}) is integrated with the same third order Runge-Kutta
scheme as the equations~(\ref{momentum}) for the velocity and magnetic
fields. Statistical data is obtained from a large number of tracer
trajectories varying from $2\cdot 10^4$ to $2\cdot 10^6$.

A detailed list of the physical parameters for the different runs is
presented in table~\ref{table2}.

\begin{table}
\centering
\footnotesize
\begin{tabular}{ccccccccccccc}
    run&$u_\perp^\mathrm{rms}$&$u_\parallel^\mathrm{rms}$&$B_\perp^\mathrm{rms}$&$B_\parallel^\mathrm{rms}$&$\epsilon_\mathrm{k}$&$\rho^{iso}_u$&$\rho^{iso}_B$&$l_K$&$\nu$  &$\tau_K$  &$L$ &$T_L$   \\
    \hline
    tgH&       1.74               & 1.28            & -                 &  -               & 2.26               & 0.74     &  -   & 0.022         &  0.008& 0.060    &1.76& 1.11      \\
    tg1&       1.50               & 0.56            &1.86               & 0.44             & 1.08               & 0.37     &0.24  & 0.027         &  0.008& 0.092    &1.88& 1.49      \\
    tg2&       1.78               & 0.63            &1.78               & 0.52             & 0.94               & 0.35     &0.29  & 0.027         &  0.008& 0.092    &3.58& 2.39      \\
    tg3&       1.87               & 0.65            &1.64               & 0.56             & 1.13               & 0.35     &0.34  & 0.026         &  0.008& 0.086    &3.43& 2.19      \\
    tg4&       1.09               & 0.43            &1.04               & 0.33             & 0.71               & 0.39     &0.31  & 0.029         &  0.008& 0.107    &1.11& 1.2       \\
    tg5&       1.92               & 0.63            &1.76               & 0.54             & 1.23               & 0.33     &0.31  & 0.03          &  0.01 & 0.092    &3.38& 2.11      \\
    ffH&       0.14              & -               & -                 & -                & $1.48\cdot10^{-3}$&$\approx 1$&$\approx 1$& 0.0195    & 0.0006 & 0.64    &2.0 & 13.8      \\
    ff &       0.13              & -               &0.066              & -                & $6.4\cdot10^{-4}$ &$\approx 1$&$\approx 1$& 0.0243    & 0.0006 & 0.99    &3.05& 24.4   
\end{tabular}
\caption{\label{table2} Parameters of the numerical
  simulations. $\epsilon_\mathrm{k}=2\nu\langle(\nabla \times
  \bm{u})^2\rangle$: mean kinetic energy dissipation rate,
  $\rho^{iso}_u=u_{\parallel}^{\rm rms}/u_{\perp}^{\rm rms}$ isotropy
  coefficient, $l_K = (\nu^3/\epsilon_\mathrm{k})^{1/4}$: Kolmogorov
  dissipation length scale, $\nu$: kinematic viscosity, $\tau_K =
  (\nu/\epsilon_\mathrm{k})^{1/2}$: Kolmogorov time scale, $L =
  (u^\mathrm{rms})^3/\epsilon_\mathrm{k}$: integral scale, $T_L =
  L/u^\mathrm{rms}$: large-eddy turnover time. For all runs:
  size of domain = 2$\pi$. For run ffH and ff only the mean rms value
  of the three components
  $u_\perp^\mathrm{rms}=u_\parallel^\mathrm{rms}=u^\mathrm{rms}$ is
  listed which differ by less than 10\%.}
\end{table}

\section{Dynamo action: structures and trajectories}
\label{sec:dynAc}
\begin{figure}[h]
%  \centering
  \includegraphics[width=0.49\columnwidth]{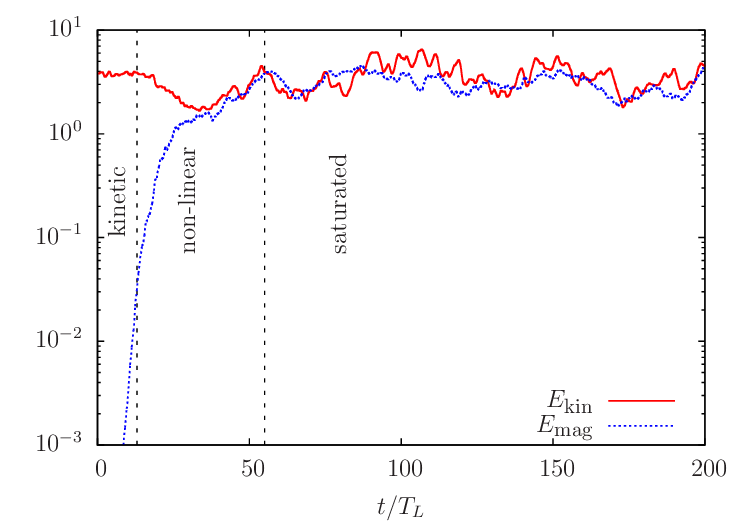}
  \hfill
  \includegraphics[width=0.49\columnwidth]{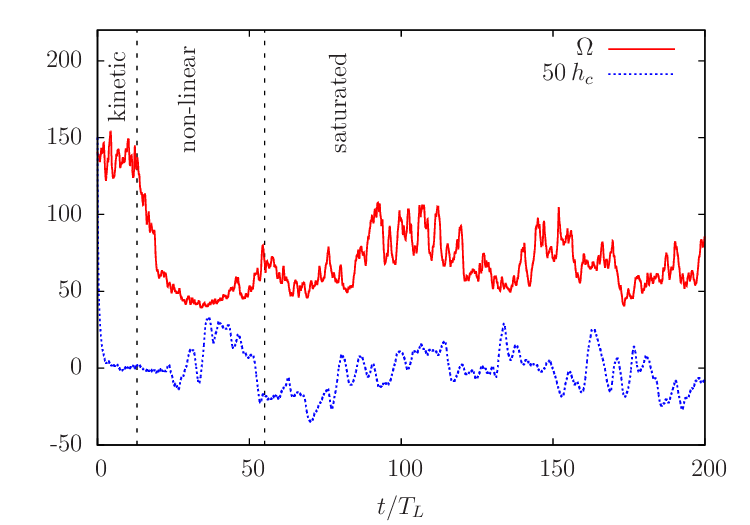}
    \caption{\label{fig:diag} Temporal evolution of kinetic and
      magnetic energy (left). In the right panel the enstrophy and
      normalized cross helicity are shown. (all data for run tg3) The
      vertical dashed lines, representing the transition between the
      different phases have been determined by using the flatness of
      the magnetic field (see section \ref{sec:magFluc} and
      Fig~\ref{fig:flatness}).}
\end{figure}

During dynamo action, three phases are clearly distinguished.  The
linear phase, where a seed of magnetic field grows exponentially by
the dynamo instability without back-reaction on the velocity field.
The second stage, called the non-linear phase, corresponds to the one
where the Lorentz force starts to act on the flow and the kinetic
energy slightly drops by the transfer of kinetic energy to the
magnetic field.  Finally the full MHD system reaches a statistically
stationary state, refereed to as the saturation regime. In
Fig.~\ref{fig:diag} (left) the kinetic and magnetic energy is
displayed for the TG flow. The three phases of the dynamo action are
separated by dashed vertical lines.
In Fig.~\ref{fig:diag} (right) the enstrophy is shown for the same
simulation.  When the magnetic energy attains a magnitude of the order
of one percent of the kinetic energy the enstrophy drops. This drop is
larger the smaller is $Pr$ (or $Rm$) that is the closer one is to the
dynamo onset. This rapid change of $\Omega$ marks the transition
between the linear phase and the non-linear phase. The saturated
regime is characterized by strong fluctuations in both energies and
enstrophy.
Note in Fig.~\ref{fig:diag} (right) that the normalized cross
helicity $h_C$ starts to fluctuate in the non-linear phase, showing a
tendency towards an alignment of the $\bm u$ and $\bm B$ fields. We
find $\langle h_C^2\rangle^{1/2}=0.03$ during the kinematic phase and
$\langle h_C^2\rangle^{1/2}=0.23$ during the saturated phase. This
alignment leads to a less efficient global electromotive force ${\bf
  u\times b}$. In the sequel we will present another quantity whose
statistics strongly changes from one regime to another and which
allows for a more precise definition of the different phases of dynamo
action. This quantity is in fact used to plot the vertical dashed
lines in Fig.~\ref{fig:diag}.

\begin{figure}[h]
  \begin{center}
    \includegraphics[width=0.49\columnwidth]{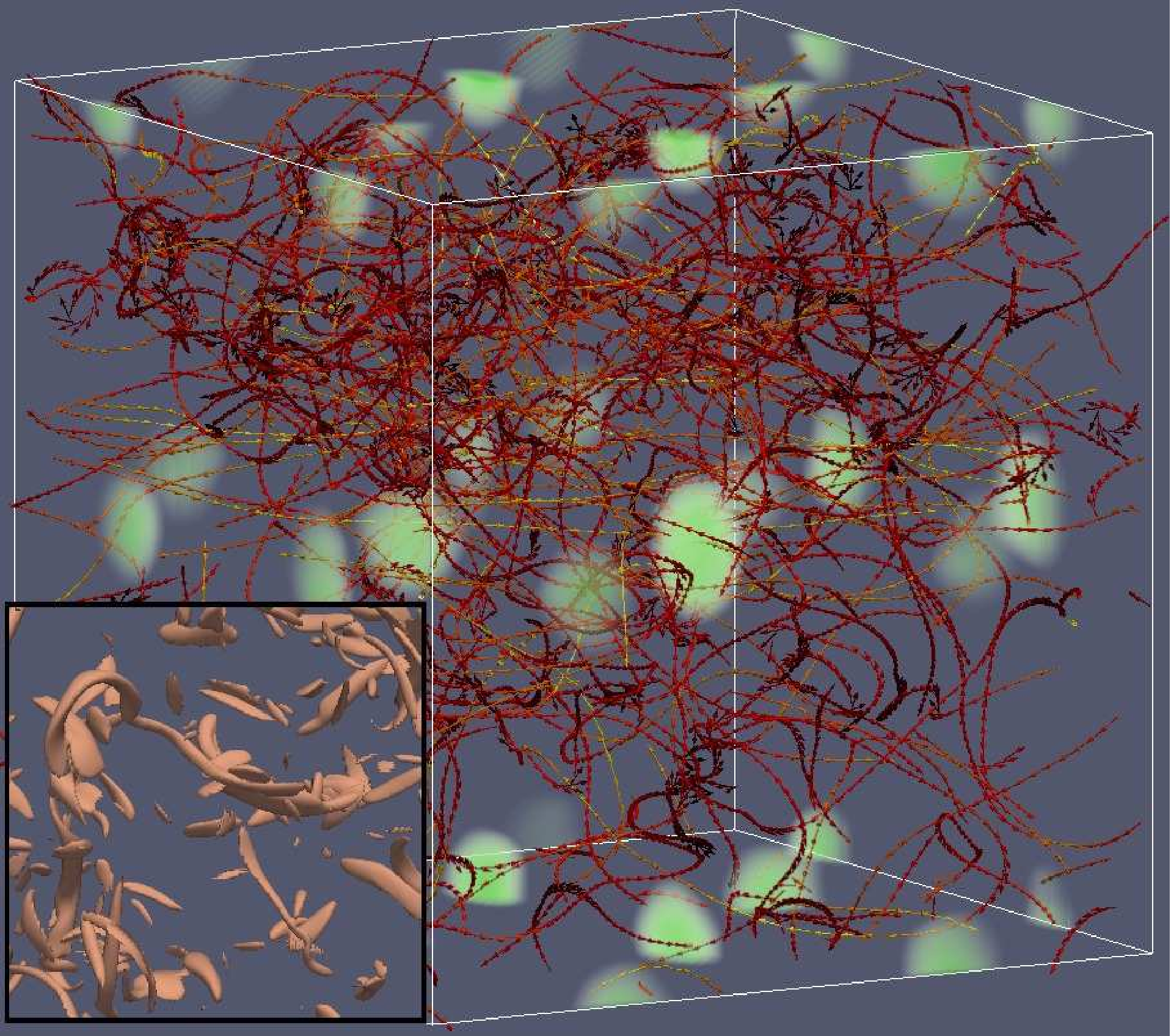}
    \includegraphics[width=0.49\columnwidth]{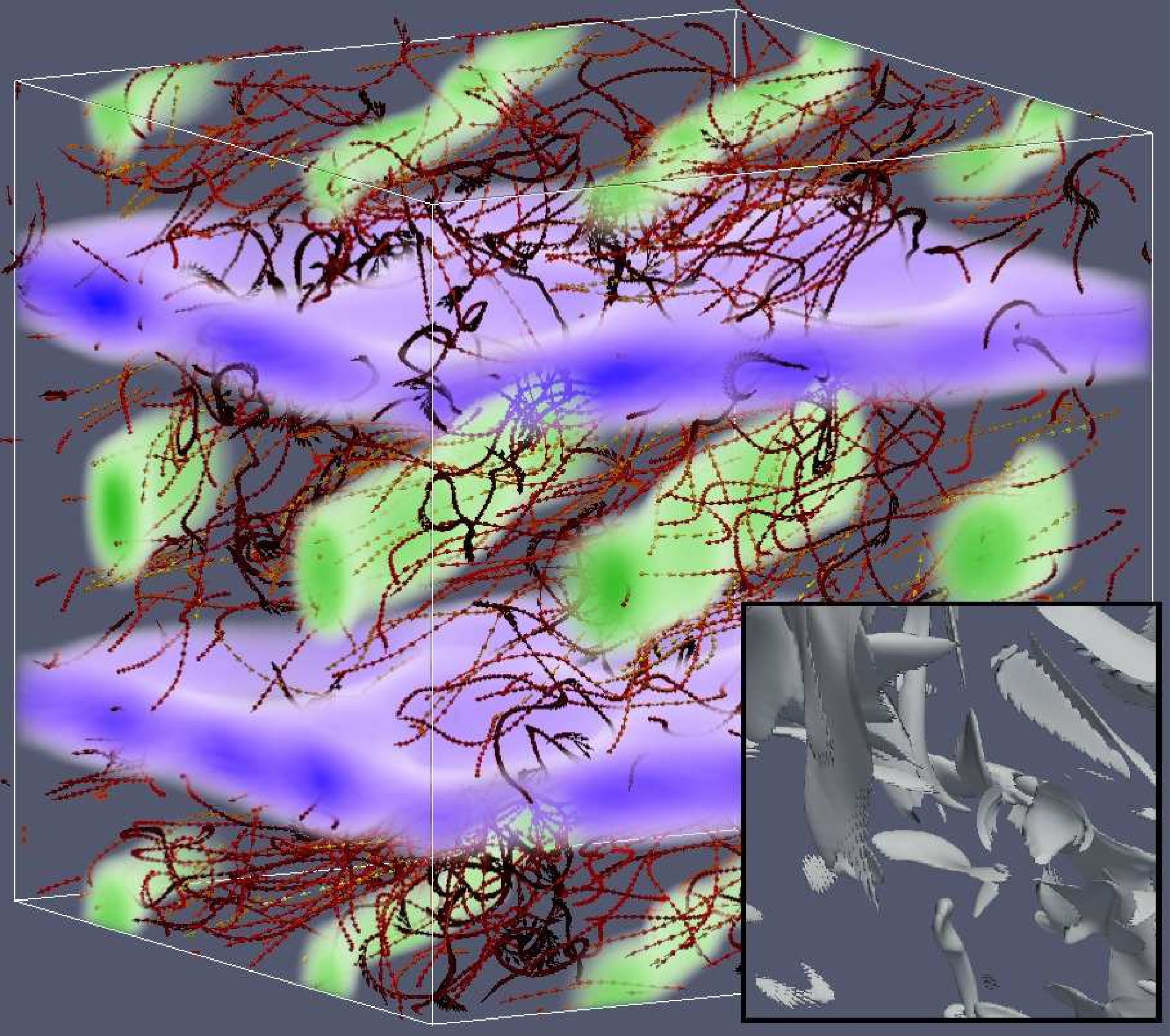}
    \includegraphics[width=0.49\columnwidth]{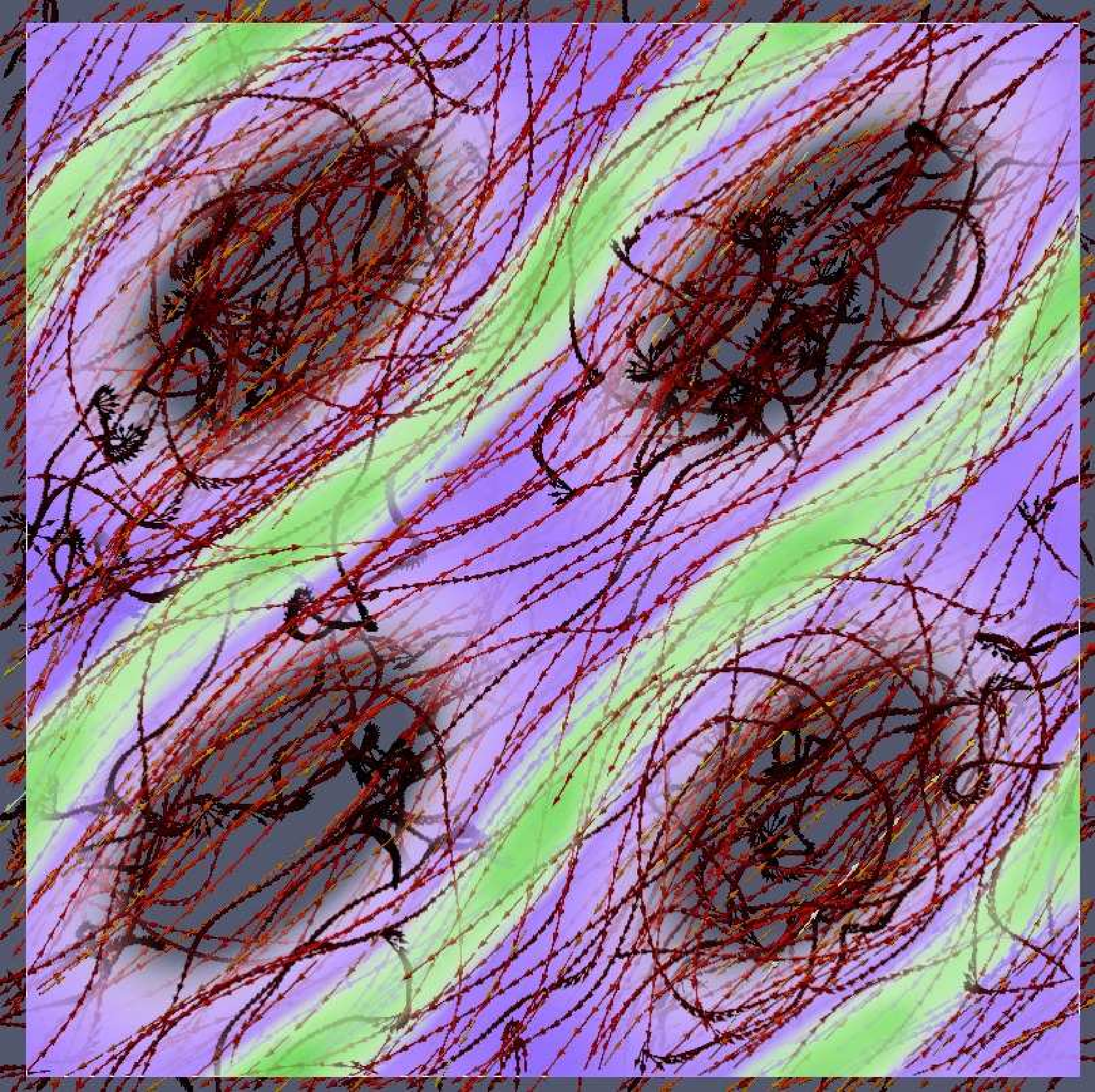}
    \includegraphics[width=0.49\columnwidth]{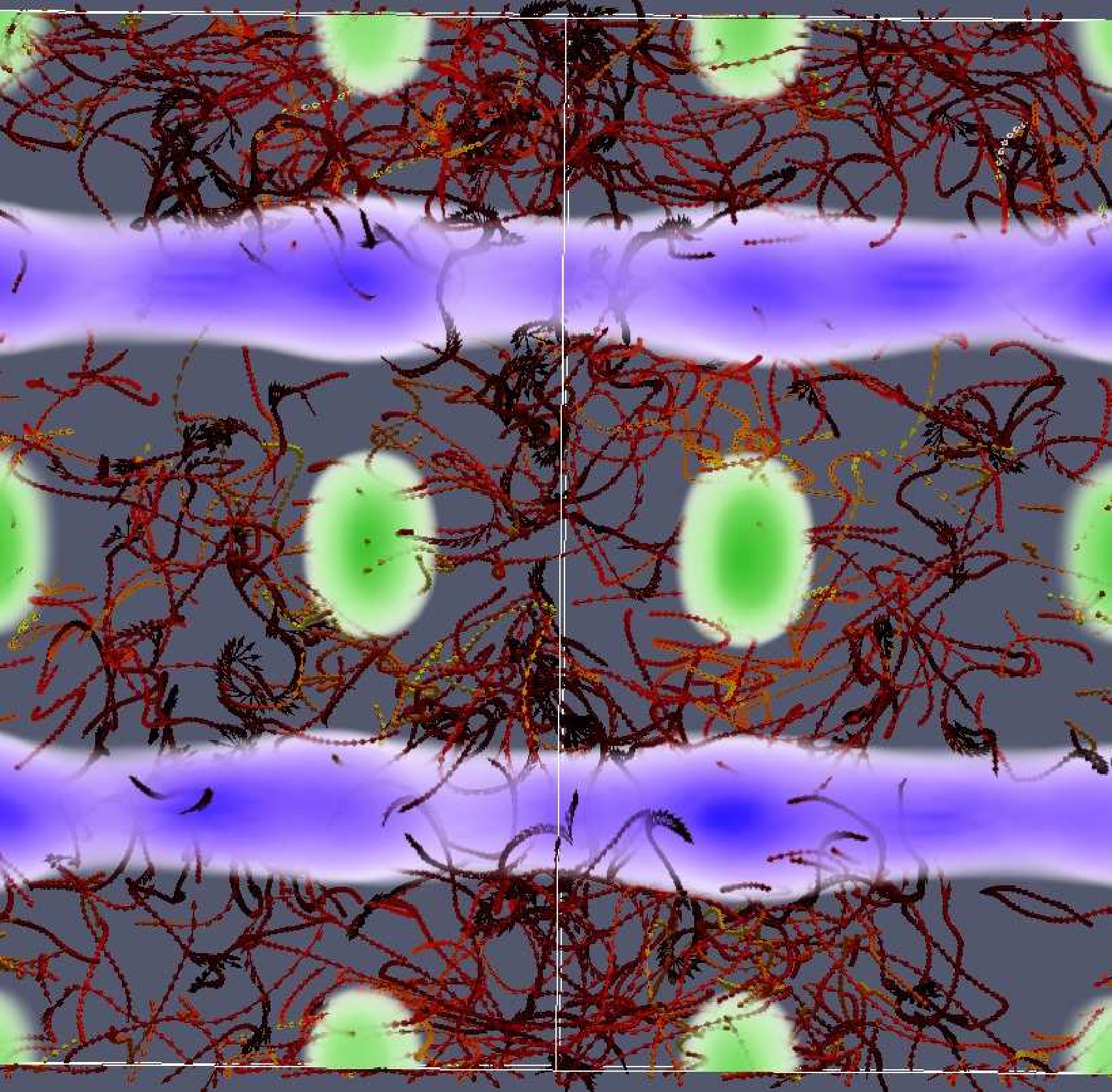}
  \end{center}
  \caption{\label{fig:traj} Volume rendering of the time-averaged
    kinetic energy (green) and magnetic energy (blue) for run tg3
    together with tracer trajectories. Their speed is given in colors
    from black (low) to yellow (high).  {\em Left top} : Linear
    phase. The inset shows a snapshot of the enstrophy, where vortex
    filaments can be observed.  {\em Right top} : Saturated phase. The
    inset shows a snapshot of the en-trophy, where MHD vortex
    filaments and sheets are visible.  {\em Left bottom} : Top
    view (along z-axis) of the saturated phase.  {\em Right bottom} :
    Transverse view (along x,y-diagonal) of the saturated phase.  }
\end{figure} 

As mentioned before, the TG flow presents a manifest anisotropy coming
from the very definition of the forcing. In the hydrodynamic regime we
find a global isotropy coefficient $\rho_u^{iso}=0.74$ (see table
\ref{table2}).  Anisotropy is even enhanced to
$\rho_u^{iso}\approx0.36$ in the saturation regime, where large scale
magnetic coherent structures appear in the two shear vortex plans and
large tubes of kinetic energy are along the diagonal direction (see
Fig.~\ref{fig:traj}). We note that for the frozen force dynamo
$\rho_u^{iso}$ is close to unity during all phases of the simulation.

%\begin{figure}[h]
%  \begin{center}
%     \includegraphics[width=0.49\columnwidth]{figures/energySpec_v}
%    \hfill
%    \includegraphics[width=0.49\columnwidth]{figures/energySpec_b}
%  \end{center}
%  \caption{\label{fig:diag} Kinetic (left) and magnetic (right) energy spectra during the linear and saturated regimes. Run \ADD{XX}.
%\NOTE{Not sure, we want to include this figure. 
%Nevertheless How do you compute the spectra? Why there is no a pick at $k=2$ for hydro case?}
%  }
%\end{figure}

We now turn to the geometry of tracer trajectories. The linear phase
is characterized by chaotic trajectories showing spiral motions around
vorticity filaments (see Fig.~\ref{fig:traj} (top left)) which
resemble trajectories in homogeneous and isotropic turbulence. The
flow is approximately homogeneous.

In the saturated phase the flow structure and the geometry of the
trajectories change (see top right panel of
Fig.~\ref{fig:traj}). Vorticity filaments of the kinematic phase are
quenched to vortex sheets in the saturated phase (see inset). The
magnetic field has an ordering and smoothing effect (see multimedia
supplementary material). Trajectories become highly aligned with the
structures of the mean (time-averaged) kinetic and magnetic
energy. These large-scale structures, well known in the literature,
interconnect the fundamental boxes of the Taylor-Green flow. In the
vicinity of the diagonally oriented high kinetic energy tubes, we
observe streams of tracers with high velocities. They are separated by
regions of chaotic motions (see Fig.~\ref{fig:traj} (bottom left))
which coincides with regions of low mean magnetic energy.  From the
bottom right panel of Fig.~\ref{fig:traj} one sees that an
approximately isotropic tracers motion of the fluid is preserved
perpendicular to the diagonal structures. For the frozen force (run
ff), there is no large scale structure traversing the fundamental box
(not shown). Blobs of increased magnetic energy are randomly
distributed all over the cube. Tracer trajectories are in this case
statistically homogeneous and isotropic.

\section{Statistics of fluctuations}
\label{sec:magFluc}
\subsection{Eulerian statistics}

\begin{figure}[h]
  \begin{center}
    \includegraphics[width=0.49\columnwidth]{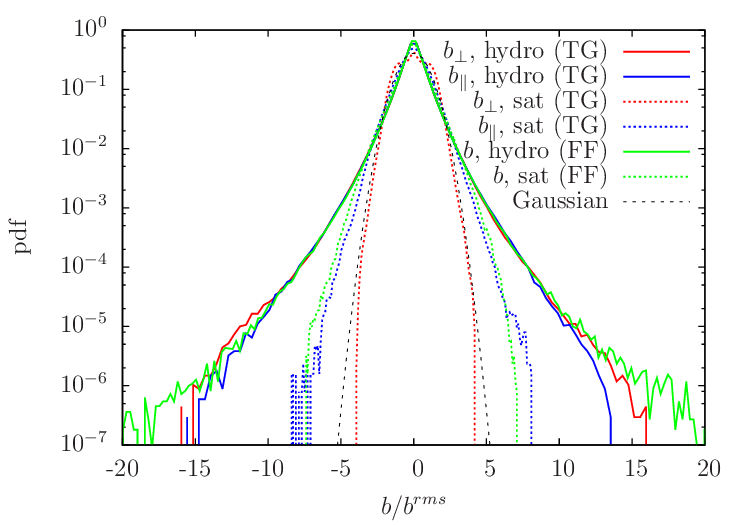} \hfill
    \includegraphics[width=0.49\columnwidth]{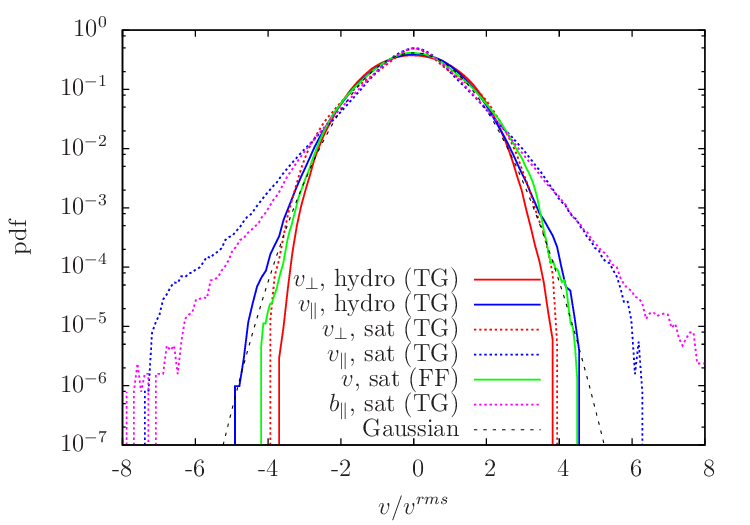}
  \end{center}
  \caption{\label{fig:eulerPDF} Probability distribution function of
    the magnetic field (left) and the velocity field (right) where the
    magnetic field component $b_\parallel, sat$ (TG) is added for
    comparison. The data of the saturated Taylor-Green regime is from
    run tg3.}
\end{figure}

A turbulent flow naturally leads to magnetic field fluctuations which
originate from stretching, twisting and diffusion of magnetic field
lines. The statistical properties of these fluctuations change
significantly from one regime to another. Especially those of the
magnetic field whose probability distribution functions (PDFs) are
shown in Fig.~\ref{fig:eulerPDF}.

In the linear phase, the PDFs of the perpendicular magnetic field
components present fat tails, far away from the Gaussian distribution
(Fig.~\ref{fig:eulerPDF} left).  Even at moderate Reynolds numbers we
observe fluctuations of the magnetic field which are 15 times larger
than its root mean square value.  We emphasize that this non-Gaussian
character of the magnetic field during the linear phase is also
observed for the frozen large scale forcing (run ffH), as apparent in
Fig.~\ref{fig:eulerPDF} (left). It has also been observed by a model
using the Recent Fluid Deformation
closure~\cite{hater-homann-etal:2011}.  This strongly suggests that
the non-Gaussianity of the growing magnetic field is a generic
property of the linear phase. In contrast, the velocity field has
nearly Gaussian statistics (see Fig.~\ref{fig:eulerPDF} right) as
usually observed in hydrodynamic turbulence.

When the flow enters the saturated phase the normalized PDFs of the
velocity and magnetic field components become similar. The tails of
the perpendicular component $b_\perp$ of the magnetic field PDF shrink
to quasi-Gaussian tails (Fig.~\ref{fig:eulerPDF} left). The
perpendicular component of the velocity field $v_\perp$ remains
Gaussian. Such magnetic-field PDFs along the saturated regime have
also been observed in the VKS experiment \cite{monchaux_von_2009}. The
parallel component of the velocity field seems to follow the magnetic
field: The PDF of $b_\parallel$ remains non-Gaussian and that of
$v_\parallel$ develops non-Gaussian tails. A reason for this matching
of velocity and magnetic field PDFs might be the increased alignment
of $v$ and $b$ in the saturated regime which couples kinetic and
magnetic fluctuations. Scale dependent alignment has been recently
evoked in a model of the scaling properties of MHD
turbulence~\cite{boldyrev:2006}.

\begin{figure}[h]
  \begin{center}
    \includegraphics[width=0.6\columnwidth]{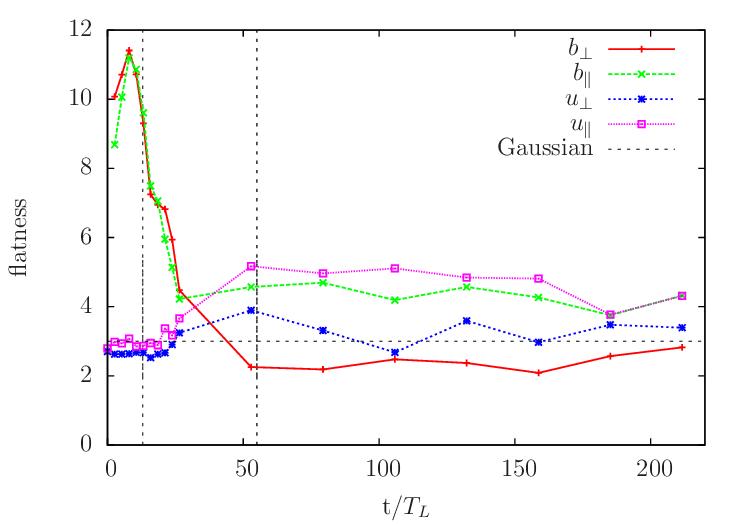}
  \end{center}
  \caption{\label{fig:flatness} Temporal evolution of the flatness of
    the magnetic and velocity field PDFs for run tg3.}
\end{figure}

In order to estimate the non-Gaussianity of a field $f$, the temporal
evolution of its flatness $\langle f^4 \rangle/\langle f^2\rangle^2$
is measured.  The flatness of the magnetic field starts from a large
value (see Fig.~\ref{fig:flatness}) originating from their stretched
tails.  During the non-linear phase its flatness strongly reduces and
reaches a slightly sub-Gaussian value for the perpendicular component
$b_\perp$ while it is super-Gaussian for the parallel component
$b_\parallel$. A significant change can also be observed for the
flatness of the velocity field PDFs (see also
Fig.~\ref{fig:flatness}). The perpendicular component $v_\perp$ starts
slightly sub-Gaussian and fluctuates around the Gaussian value $3$,
while the flatness of the component $v_\parallel$ is clearly larger
and comparable to that of $b_\parallel$ during the saturated phase.
These changes in the behavior of the flatness temporal evolution allow
to clearly distinguish the three phases of the dynamo action indicated
by the vertical lines drawn in Fig.~\ref{fig:diag} and
\ref{fig:flatness}).

\begin{figure}[h]
  \begin{center}
    \includegraphics[width=0.49\columnwidth]{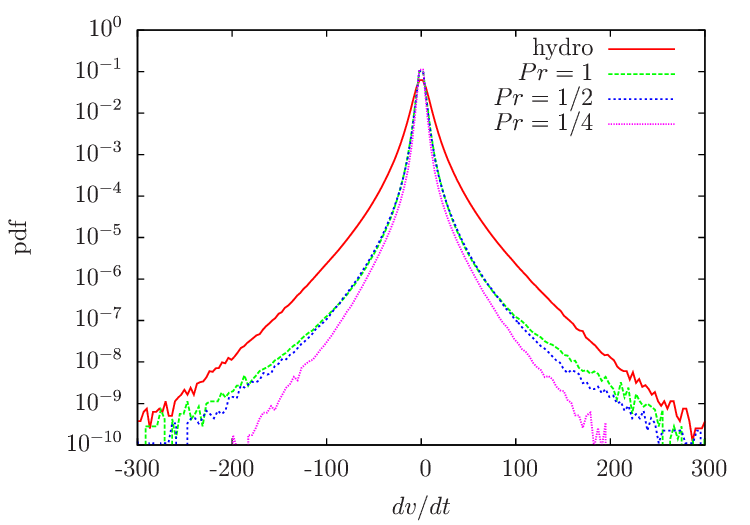} \hfill
    \includegraphics[width=0.49\columnwidth]{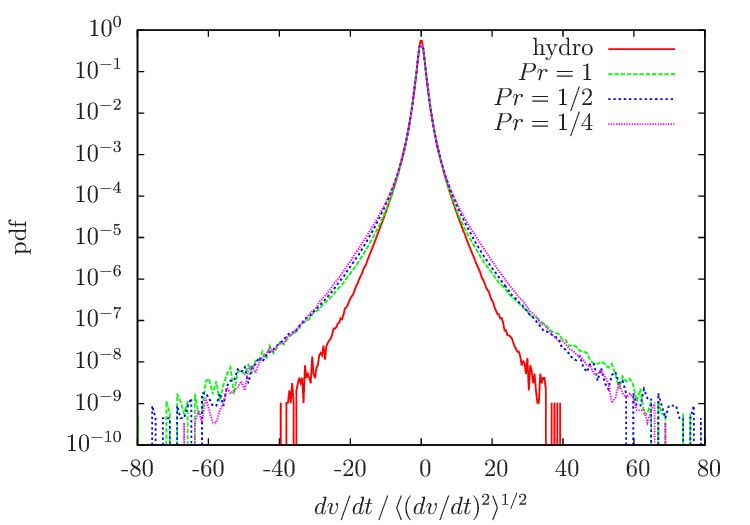}
  \end{center}
  \caption{\label{fig:PDF_acc} Probability distribution function of
    acceleration. Non-normalized (left) and normalized to standard
    deviation (right) for a hydrodynamic simulation (run tgH) and
    saturated regimes (run tg1, tg2, and tg3).}
\end{figure}

We now turn to the statistics of a small scale quantity of the flow,
namely the fluid acceleration, whose PDFs are displayed in
Fig.~\ref{fig:PDF_acc} (left). Contrarily to the PDFs of the velocity,
a large scale quantity, there are no significant differences among
perpendicular and parallel components. This is agreement with the
general observation that anisotropy of large scales decreases towards
the smaller scales in turbulent flows. In order to improve statistics,
the three acceleration components have been averaged.  The
acceleration has strongly non-Gaussian tails as usually observed in
turbulent flows \cite{porta-bodenschatz-etal:2001} and decrease from
the kinematic to the saturated phase (see Fig.~\ref{fig:PDF_acc}
(left)).  This is an agreement with the known fact that the saturation
regime smoothes out the small scales of the velocity.  Indeed, the rms
acceleration reduces from $a_{rms}=80$ for the hydrodynamic phase to
$a_{rms}=16$ and $a_{rms}=10$ for the saturated phase for $Pm=1$ and
$Pm=1/4$, respectively. Smaller magnetic Prandtl numbers imply thus
smaller accelerations. However, when comparing the normalized PDFs
(normalized to standard deviation, Fig.~\ref{fig:PDF_acc} (right)),
more extreme accelerations events are observed for the saturated
regime than the pure hydrodynamic case.  The tails of the PDFs become
significantly fatter. The flatness of the PDFs reaches $25$ in the
saturated regime, while it is $14$ in the hydro case. Here, the
Prandtl number has a negligible effect. The normalized PDFs
distinguish therefore clearly the saturated from the kinematic
dynamics.

\subsection{Lagrangian statistics}

We now turn to study the properties of the magnetic field along fluid
element trajectories, i.e. the magnetic field seen by a tracer. We
will first focus on changes in the perpendicular component of the
magnetic field, the dominant component of the mean field. For this, we
consider the temporal increment $\delta b_\perp(\tau,t,\bm{x}) =
B_\perp(\bm{X}(\bm{x},t+\tau),t+\tau)-B_\perp(\bm{X}(\bm{x},t),t)$ where
$B_\perp$ denotes a magnetic field component from the perpendicular
plane ($B_x$ or $B_y$) and $\bm{X}(\bm{x},t)$ a fluid trajectory starting
at $\bm{x}$ at a time $t$. To improve statistics we average over
$\bm{x}$ and $t$ which leads to the study of the magnetic field
increment
\begin{equation}
\label{deltaB}
\delta b_\perp(\tau) = b_\perp(\tau) - b_\perp(0).
\end{equation}
%
%We note that we exclude the discussion of the kinematic phase as the
%system is not in statistical equilibrium.

\begin{figure}[t]
  \begin{center}
    \includegraphics[width=0.49\columnwidth]{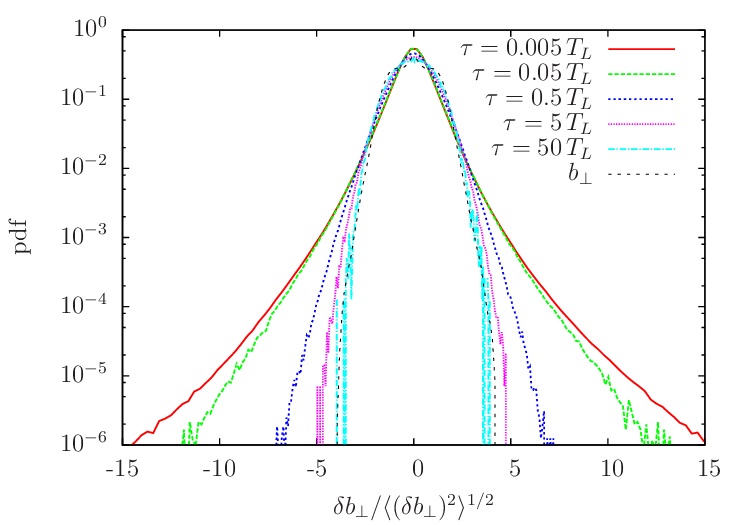} 
    \includegraphics[width=0.49\columnwidth]{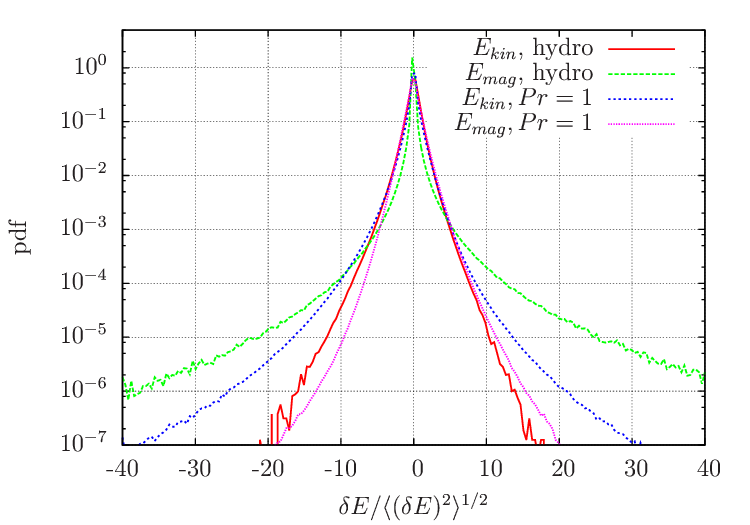}
  \end{center}
  \caption{\label{fig:lagPdfB} {\em Left}: PDFs of the magnetic field
    increments in the perpendicular direction along fluid element
    trajectories for several time lags in the saturated phase. For
    comparison the PDF of the $b_\perp$ is shown. (data from run tg3)
    {\em Right}: PDFs of the magnetic field energy increments for
    $\tau$ of the order of $5\cdot 10^{-3}T_L\sim \tau_K$. (data
    from run tg3)}
\end{figure}

The normalized PDFs of the increment $\delta b_\perp$ during the
saturated phase are shown in Fig.~\ref{fig:lagPdfB} (left). Stretched
tails were also observed for small scale magnetic fluctuations in the
solar wind~\cite{alexandrova}. With increasing time-lag $\tau$ the
PDFs show a transition to Gaussian statistics for large time
lags. This is simply due to time decorrelation of $b_\perp(\tau)$ and
$b_\perp(0)$ so that one recovers for large $\tau$ the one-point
Eulerian PDF shown in Fig.~\ref{fig:eulerPDF} (left). It is
important to stress that the time scale $\tau$ at which this
decorrelation happens is unusually long as it exceeds significantly
one large-eddy turn-over time. The associated long time regime will be
analyzed in detail in the subsequent section.

Another type of Lagrangian increment has recently attracted interest
in the hope of a better understanding of the energy cascade and energy
flux in turbulent flows~\cite{xu-pumir-etal}, namely the kinetic (and
magnetic) energy increment
\begin{equation}
  \label{deltaB2}
  \delta E_v(\tau) = \frac{1}{2}(\bm{v}^2(\tau) - \bm{v}^2(0)).
\end{equation}
In Fig.~\ref{fig:lagPdfB} (right) the corresponding PDFs are shown for
a short time lag $\tau$. The mean of all PDFs but that of the magnetic
field energy during the linear phase is zero. Only the mean of the
latter is positive because of the exponential increase of magnetic
energy. More interesting is their skewness $S_f=\langle f^3
\rangle/\langle f^2\rangle^{3/2}$. For the hydro case the PDF of the
kinetic energy has a negative skewness of $S_v\approx -0.4$. We
remark that the Lagrangian derivative $d |\bm u|^2/dt$ can be written
in terms of Eulerian quantities $d |\bm u|^2/dt=2(|\bm u|^2\nabla_{\bm
  u}^{\parallel}\bm{u}+\bm{u}\partial_t \bm{u})$, where $\nabla_{\bm
  u}^{\parallel}\bm{u}$ is the longitudinal velocity gradient
$\nabla^{\parallel}_r\bm{u}\equiv \hat{\bm r} \cdot (\hat{\bm r} \cdot
\nabla \bm u)$ evaluated in the direction of the local velocity
$\hat{\bm u}$. $\nabla^{\parallel}_r\bm{u}$ averaged over $\hat{\bm
  r}$ is known to have negative skewness close to the one measured for
$d |\bm u|^2/dt$. A negative skewness means that violent decelerations
are more probable than violent accelerations. The skewness of the
velocity PDFs decreases towards the saturated regime
($S_v\approx-1.0$ for $Pm=1/4$, $S_v\approx-1.2$ for $Pm=1/2$ and
$S_v\approx-1.2$ for $Pm=1$).  In contrast, the magnetic field PDF
has a positive skewness $S_b\approx 2.5$ during the kinematic phase,
which decreases in the saturated phase ($S_b\approx 0.43$ for
$Pm=1/4$, $S_b\approx 0.32$ for $Pm=1/2$ and $S_b\approx 0.34$ for
$Pm=1$).

The evolution of the energy increment skewness along the different
dynamo regimes, is related to the energy exchange between the kinetic
and the magnetic energies in the non-linear and saturation
regimes. Note that energy transfers are non-linear exchanges, with
local and non local transfer in a large range of scale
\cite{alexakis_shellshell_2005, mininni_shellshell_2005}.

%% Even during the saturated phase when the magnetic energy is constant
%% up to fluctuations we still observe increases and decreases of the
%% magnetic field at all time scales. In order to better understand where
%% the generation and destruction of the magnetic field happens we
%% visualize in Fig.~\ref{fig:bChanges} time-averaged changes of the
%% magnetic field strength $\delta B(0.1\,\tau_\eta)=\langle
%% |\bm{B}(\bm{x},t+0.1\,\tau_\eta)|-|\bm{B}(\bm{x},t)|\rangle$.
%% \begin{figure}[h]
%%  \begin{center}
%%    \includegraphics[width=0.45\columnwidth]{figures/b_db}
%%    \hfill
%%    \includegraphics[width=0.52\columnwidth]{figures/dbZ}
%%  \end{center}
%%  \caption{\label{fig:bChanges} Left: Time averaged changes $\delta
%%    B(0.1\,\tau_\eta)$ of the magnetic field along tracer
%%    trajectories. Red: increases, Blue: decreases. Right. Averaged
%%    data of the left panel along the $z$ axis.}
%% \end{figure}
%% On average one observes increases on a regular 3x3x4 grid which
%% enclosed the regions of the mean magnetic field energy but does not
%% coincide with them. In the centers of the highest mean magnetic energy
%% on average neither increases nor decreases happen. Both are in
%% quasi-equilibrium. The diminution of magnetic field happens on
%% average in the regions with weak mean magnetic field.

\section{Long time scaling regime}
\label{sec:longTime}

\begin{figure}[h]
  \begin{center}
    \includegraphics[width=0.49\columnwidth]{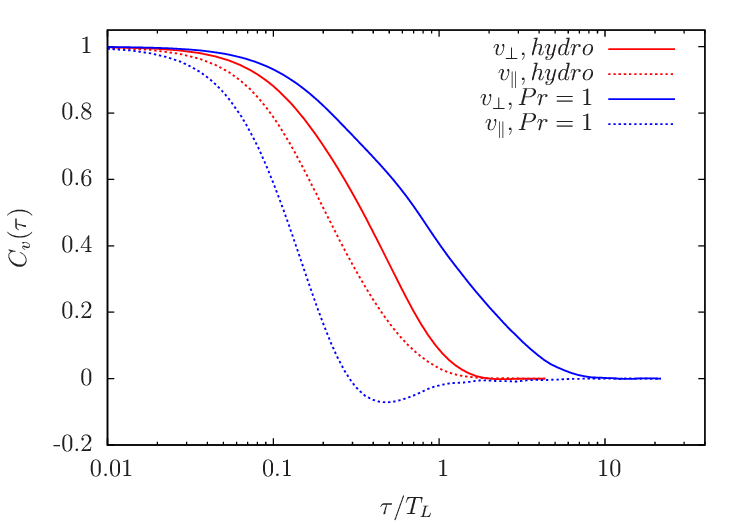}
    \includegraphics[width=0.49\columnwidth]{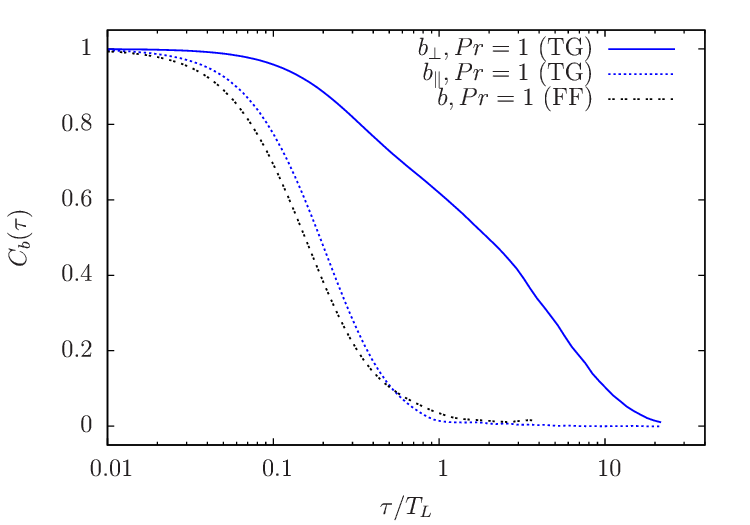} 
  \end{center}
  \caption{\label{fig:lagCorr} {\em Left}: Autocorrelation functions
    $C_v(\tau)$ of the velocity along trajectories in the hydrodynamic
    phase (run tgH) and saturated phase (run tg3) {\em
      Right}: Autocorrelation functions $C_b(\tau)$ of the magnetic
    field measured along trajectories for the saturated phase for run
    tg3, and ff.}
\end{figure}

In the previous section we observed that the transition to Gaussian
statistics of the perpendicular magnetic field takes very long times
during the saturated regime while this happens on time scales of the
order of $T_L$ in the kinematic regime. The reason are long lasting
correlations. The autocorrelation of the Lagrangian velocity
increments $C_v(\tau) = \langle v_i'(t) v_i'(t+\tau) \rangle / \langle
v_i'^2(t)\rangle$ with $v_i'(t)= v_i(t) - \langle v_i(t) \rangle$, is
displayed in Fig.~\ref{fig:lagCorr} (left). The Lagrangian correlation
time is indeed of the order $T_L$ for the hydrodynamic regime. During
the saturated regime the perpendicular component of the velocity field
remains correlated for a longer time (see Fig.~\ref{fig:lagCorr})
(left)) of the order of several $T_L$. Note that the parallel
component becomes anti-correlated at times of the order of
$\pi/v^{rms}_z$ (see table \ref{table2}). This corresponds to the
typical time of a fluid particle crossing a TG fundamental box in the
$z$ direction and starting to feel the mirror symmetries of the TG
flow. 

The long correlation time scale is also clearly observed for the
magnetic increments in Fig.~\ref{fig:lagCorr} (right) and corresponds
to the time-lag for which the PDF of the magnetic field increments
becomes Gaussian. We note that this time scale is a specific property
of the TG flow which is not observed with the frozen force (run ff,
Fig.~\ref{fig:lagPdfB} (right)) and it does only weakly depend on
$Pm$.

In order to better understand the long time correlations and the
non-Gaussianity of the magnetic field increments we study Lagrangian
structure functions (LSF) of the form
\begin{equation}
  S^B_p(\tau)=\langle |B_\perp(t+\tau)-B_\perp(t)|^p \rangle
\end{equation} 
during the saturated phase. Here again, $B_\perp$ denotes a magnetic
field component from the perpendicular plane ($B_x$ or $B_y$). In
turbulent flows structure functions are usually measured to identify
and to analyze the scales of the inertial range (for $u_i$ instead of
$B_i$) in which they behave as a power law $S_p(\tau)\sim
\tau^{\zeta}_p$. The corresponding scaling exponents $\zeta_p$
provide a handle on the phenomenon of intermittency which
characterizes the occurrence of extreme events in the dynamical
evolution of the system~\cite{frisch:1995}.

\begin{figure}[h]
  \begin{center}
    \includegraphics[width=0.65\columnwidth]{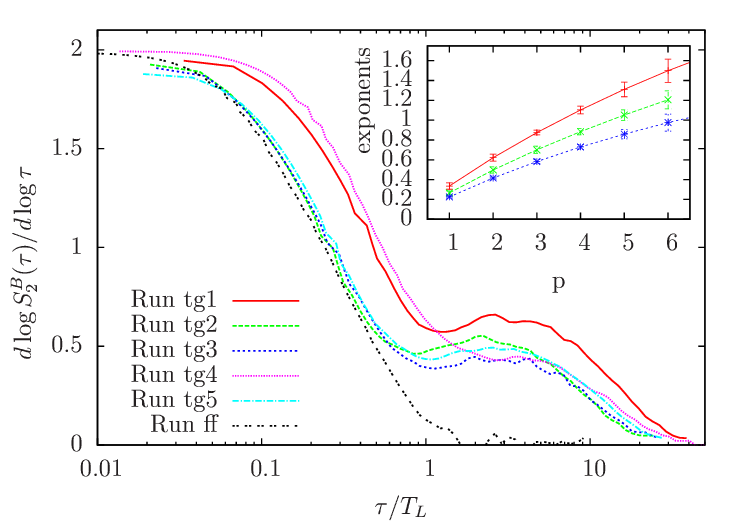} \hfill
  \end{center}
  \caption{\label{fig:LSF} Local slope of the second order Lagrangian
    structure functions of the magnetic field $S^B_2(\tau)$ for all
    runs in the saturated regime. The inset shows the measured scaling
    exponents $\zeta^B_p$ different orders $p$. Error bars give the
    maximal variation around the mean in the interval where $S_2^B$ is
    flat. (data from run tg1, tg2, and tg3)}
\end{figure}

A scaling range and corresponding scaling exponents can reliably be
measured via the local slope of the structure functions that is their
logarithmic derivative $d\log S_p / d\log \tau$. This quantity for the
second order LSF $S^B_2$ is shown in Fig~\ref{fig:LSF} for all
runs. In hydrodynamic turbulence, even for very large Reynolds
numbers, the range of scales where a power law is observed and where
thus the local slope is constant is very narrow if visible at
all~\cite{falkovich-pumir-etal-2012}. We therefore do not expect to
observe a power-law behavior on time-scales shorter than
$1\,T_L$. Surprisingly, for larger times scales we observe a clear
plateau for the Taylor-Green runs. Its width is similar (from
approximately $1\,T_L$ to $10\,T_L$) for all runs but run tg4 (forced
at $k_0=2$) indicating a relationship with the forcing.  The
corresponding scaling exponent does not seem to depend on the Reynolds
number of the flow while it changes depending on the magnetic Prandtl
number: It decreases with increasing $Pm$.  As expected, for the
frozen-force simulation (ff) where correlations do not persist beyond
$1\,T_L$ no scaling region is observed.

Higher order statistics reveal that the magnetic fluctuations in this
scaling region are intermittent. The scaling exponents $\zeta^B_p$ of
the structure functions $S^B_p$ up to order six are shown in the inset
of Fig.~\ref{fig:LSF} for three different Prandtl numbers. They reveal
that the long time magnetic fluctuation are intermittent as all curves
are bent. Concerning their magnetic Prandtl number dependence one
finds that the values of the exponents grow with decreasing $Pm$ but
we note that the curves of the relative scaling exponents
$\zeta_p/\zeta_2$ fall on top of each other within the error bars (not
shown).

In order to better understand the origin of the observed long time
correlations we computed the auto-correlations of the terms on the
right hand side of (\ref{eq:mhd1}) multiplied by $\bm u$. They
corresponds to the correlation of the energy sources and sinks of the
velocity field. We find that the correlation time of the mean external
energy injection rate $\langle \bm u\cdot \bm F\rangle$ becomes longer
during the non-linear phase and extends well beyond one $T_L$ during
the saturated phase. It is then also on this time scale that the total
energy of the system fluctuates. This is in agreement with the
observed changes of the correlation time of $u_\perp$ in
Fig.~\ref{fig:lagCorr} (left) as the Taylor-Green force is
constant. The energy exchange time scale $\langle \bm u\cdot
(\nabla\times{\bm B}) \times {\bm B} \rangle$ connected to the Lorentz
force is shorter. We note that the mean external energy injection rate
is a large-scale quantity. Apparently, the Lorentz force changes the
global flow structure in such a way that the resulting large-scale
velocity field and the (constant) Taylor Green force remain correlated
for times much longer than without Lorentz force. The observed scaling
regime is probably not a classical turbulent scaling regime.

\begin{figure}[t]
  \begin{center}
    \includegraphics[width=0.49\columnwidth]{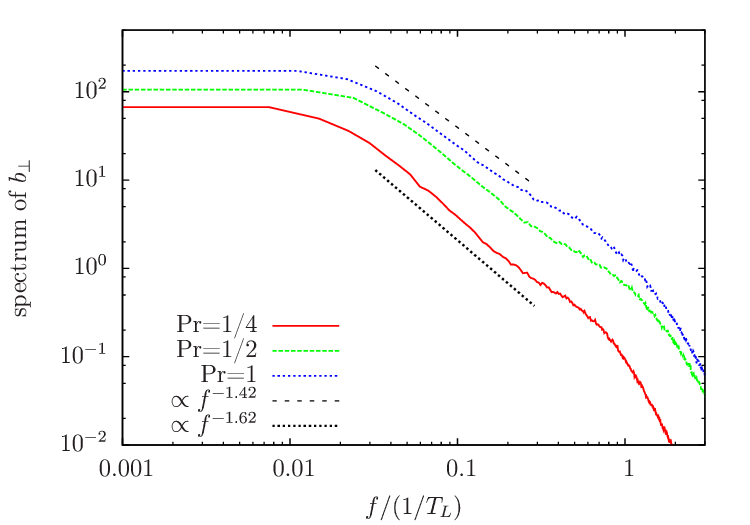}
    \hfill
    \includegraphics[width=0.49\columnwidth]{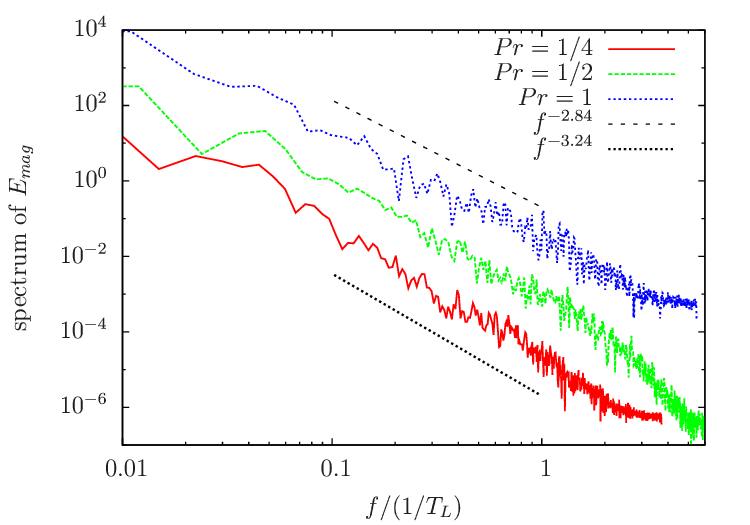}
  \end{center}
  \caption{\label{fig:spec} Left: Power density spectrum (PDS) of
    $b_\perp$ for run tg1, tg2, and tg3. The frequency $f$ is given in
    terms of the large scale frequency $1/T_L$. Right: PDS of the
    total magnetic energy $E_{\rm mag}(t)$ for run tg1, tg2, and
    tg3. All curves are shifted for clarity.}
\end{figure}
We complement the discussion of the long time scaling regime by
analyzing the power density spectrum (PDS) of the perpendicular
magnetic field $b_\perp$ along tracer trajectories. For the PDS we
find a low frequency band showing a power-law decay (see
Fig.~\ref{fig:spec} (left)). The slope $\zeta_2$ of a second order
structure function and that of the corresponding PDS ($\alpha$) are
connected via the formula $\alpha=-\zeta_2-1$~\cite{frisch:1995}. The
measured slopes of the PDS are consistent with the scaling exponents of
the structure functions.

Finally, we would like to note that we measured slopes of the spectrum
of the temporal fluctuations of the total magnetic energy $E_{\rm
  mag}(t)$ (see Fig.~\ref{fig:spec} (right)) which are compatible with
the previously presented data. Keeping in mind that the variable of
measured PDS has a dimension of $B^2$ and not of $B$ like that of the
LSF we find an agreement with $\zeta^L_2$ by simple dimensional
arguments.

\section{Conclusions}
\label{sec:conclusion}

By means of direct numerical simulations of the magnetohydrodynamic
equations forced by the Taylor-Green vortex we analyzed Lagrangian
statistics during the three stages of a dynamo system: the kinematic
phase with negligible magnetic field energy, the non-linear phase with
awakening Lorentz force, and the saturated phase when the non-linear
dynamics lead to a statistically stationary equilibrium.

Lagrangian data allows for a clear identification of these three
regimes. The statistics of the magnetic and the velocity field changes
drastically from one phase to another. The magnetic field PDFs are
strongly non-Gaussian in the kinematic phase and quasi-Gaussian in the
saturated regime. PDFs of magnetic and velocity energy increments
along tracer trajectories are skewed. We find a positive skewness for
the magnetic energy which reduces from the kinematic to the saturated
phase. The skewness of the kinetic energy is negative during the
kinematic phase and is even smaller during the saturated phase. The
evolution of the skewness of PDF energy increment is clearly due the
energy transfer between the kinetic and magnetic energies.

Temporal correlations extend much longer in the saturated regime than
in the kinematic one and exceed significantly one large-eddy turn-over
time. This is in agreement with increased temporal correlations of the
external kinetic energy injection rate due to the action of the
Lorentz force. Remarkably, we find a clear scaling regime of magnetic
field increments along particle trajectories at time scales
approximately ranging from one to ten large-eddy turn-over
times. High-order scaling exponents of the Lagrangian structure
functions show that these long-time magnetic field fluctuations are
intermittent and that intermittency is increasing with increasing
magnetic Prandtl number. The second-order scaling is consistent with a
power-law observed for the corresponding frequency spectrum and that
of fluctuations of the total magnetic energy of the system.

Finally, we would like to briefly mention that we do not observe a
long time scaling regime for simulations with an ABC force. If it is
specific to the Taylor-Green force or if also other forces develop it
has to be investigated in future work.
\subsection*{Acknowledgments}
We thank J.F. Pinton for useful discussions. Access to the IBM
BlueGene/P computer JUGENE at the FZ J\"ulich was made available
through project HBO22 and partly through project HBO36. Computer time
was also provided by GENCI and the Mesocentre SIGAMM machine. This
work benefited from support through DFG-FOR1048 \\

\bibliographystyle{unsrt}
%\bibliographystyle{ieeetr}
%\bibliographystyle{aipnum4-1}
%\bibliography{bib}

\end{document}